\documentclass[a4paper,UKenglish,cleveref]{lipics-v2021}
\usepackage{listings}
\usepackage{booktabs}
\usepackage{amssymb}  

\nolinenumbers

\title{Carnap Ten Years Later:}
\subtitle{Lessons Learned and Next Steps}
\titlerunning{Carnap Ten Years Later}

\author{Graham Leach-Krouse}{Draper Laboratory, USA}{gleachkr@gmail.com}
  {https://orcid.org/0000-0000-0000-0000}{}

\authorrunning{G.~Leach-Krouse}
\Copyright{Graham Leach-Krouse}



\category{}
\relatedversion{}
\acknowledgements{}

\EventEditors{Editor One and Editor Two}
\EventNoEds{2}
\EventLongTitle{Conference Name}
\EventShortTitle{Conference Acronym}
\EventAcronym{CONF}
\EventYear{2025}
\EventDate{Month 1--3, 2025}
\EventLocation{City, Country}
\EventLogo{}
\SeriesVolume{1}
\ArticleNo{1}

\begin{document}

\maketitle

\begin{abstract}
  The first part of this paper provides an experience report, recounting the
  design and long-term maintenance of the Carnap proof assistant framework used
  cumulatively by over 45,000 students worldwide over the last decade. We cover
  the good, the bad, and the ugly: what worked well, what didn't work, and what
  added friction to development and maintenance over time.

  These insights motivate a bottom-up redesign of the Carnap framework, which is
  the topic of the paper's second part. Briefly, the new design combines
  a new high-performance verifier kernel (\texttt{mm0-zig}) targeting Mario
  Carneiro's metamath zero format, and the Aufbau Bytecode Compiler,
  (\texttt{abc}), a new proof compiler that can serve as a backend for richly
  interactive proof-authoring experiences on the web.
\end{abstract}

\section{Introduction}

In 2015, I was teaching three sections of formal logic at Kansas State
University. Because I am a sinner, I could not find the necessary strength
within myself to grade the hundreds of pages of natural deduction and truth
tables that students needed graded. But because I am not entirely a wretch,
I couldn't bring myself to lower my standards for student engagement.

Elementary logic is in the fingers as much as it is in the brain. It requires
a habit of thought (really many habits) that can only be arrived at through
practice. Eventually students might be lucky enough to reach escape velocity,
go supercritical, or whatever the right metaphor is for reasoning formally
under your own power. But until they do, it's the instructor's responsibility
to keep feeding energy into the system. Energy, in this case, comes from the
friction a student experiences when they confront an abstract puzzle that does
not admit a simple or approximate solution. And they will only know that the
puzzle can't be solved simply, or approximately, if they do the problem sets,
and the problem sets are carefully and accurately marked.

My first attempt to extricate myself from the tight pinch I was in---caught
between my pedagogical ideals and my slothful nature---was to do what I had seen
older, wiser professors do. I tried out a bit of teaching software, David
Kaplan's Logic 2010. It had many excellent features: it was free, it was based
on Kalish and Montague\footnote{\emph{Logic: Techniques of Formal Reasoning}
  \cite{kalishmontague}, an
  unusually rich and rigorous introductory logic text. If Kalish and Montague
  was too hard, you could also fall back to Terence Parsons' \emph{An Exposition
  of Symbolic Logic} \cite{parsonssymboliclogic} (``the terry text''), which was a little friendlier and
  also Logic 2010 compatible.
}, and it worked hard to tie logic to the analysis of ordinary language in a way
that showed some of the richness of the subject beyond the undeniable elemental
joy of symbol pushing. In the end though, I couldn't make it work. Students
found the installation process onerous, and were impatient with the dated user
interface.

Flashier tools were available, but they cost money, and I really \emph{liked}
Logic 2010. So I made what turned out to be a fairly consequential decision:
I decided to build my own tool. It needed to be:
\begin{enumerate}
  \item Web based---to avoid the initial three weeks or so of nagging my
    students to install the software properly.
  \item Free---many of my students didn't have a lot of money, and they didn't
    need another expensive textbook to pay for.
  \item Engaging---insofar as possible, it was important for me to try to get
    my students into a flow state where they could be genuinely absorbed
    in a problem.
  \item Flexible---I had grand ambitions to use the tool to teach modal logic
    and other atypical subjects. I also just had a general hankering for
    absolute generality, a bit of a professional hazard as a philosopher.
\end{enumerate}
The result was \url{https://carnap.io}.\footnote{
  The name was a tribute to Rudolf Carnap, who seemed like a good figurehead
  for a tool that was intended to support wide variety of logical formalisms.
  I also found out recently that David Kaplan was Rudolf Carnap's last graduate
  student. I do not know what to make of this strange coincidence.
} It started as a small platform, which I used only in my own teaching. But at
a certain point I had invested enough work that it seemed silly not to share it.
I posted the result publicly, got some uptake, and continued work. During the
Covid 19 pandemic, usage increased dramatically, and I did my best to keep up
with demand by adding support for dozens of new deductive systems, and scaling
the primitive server backend to keep up with heavier workloads, learning as
I went. By 2026, Carnap has graded over four million problems, and served over
45,000 students.

The first part of this paper is about what worked, what didn't, and what I wish
I had done differently. The second part of the paper introduces two new pieces
of technology motivated by some of the lessons learned during my time with
Carnap. The first of these is the Aufbau Bytecode Compiler, a flexible proof
compiler which produces independently verifiable binary proof certificates in
the Metamath Zero binary format. The second is mm0-zig, a small trusted kernel
that verifies MMB certificates. Both tools compile to WASM and run in the
browser.

\section{My Time with Carnap}

\subsection{Carnap at a Glance}

\begin{figure}
  \centering
  \includegraphics[width=.85\linewidth]{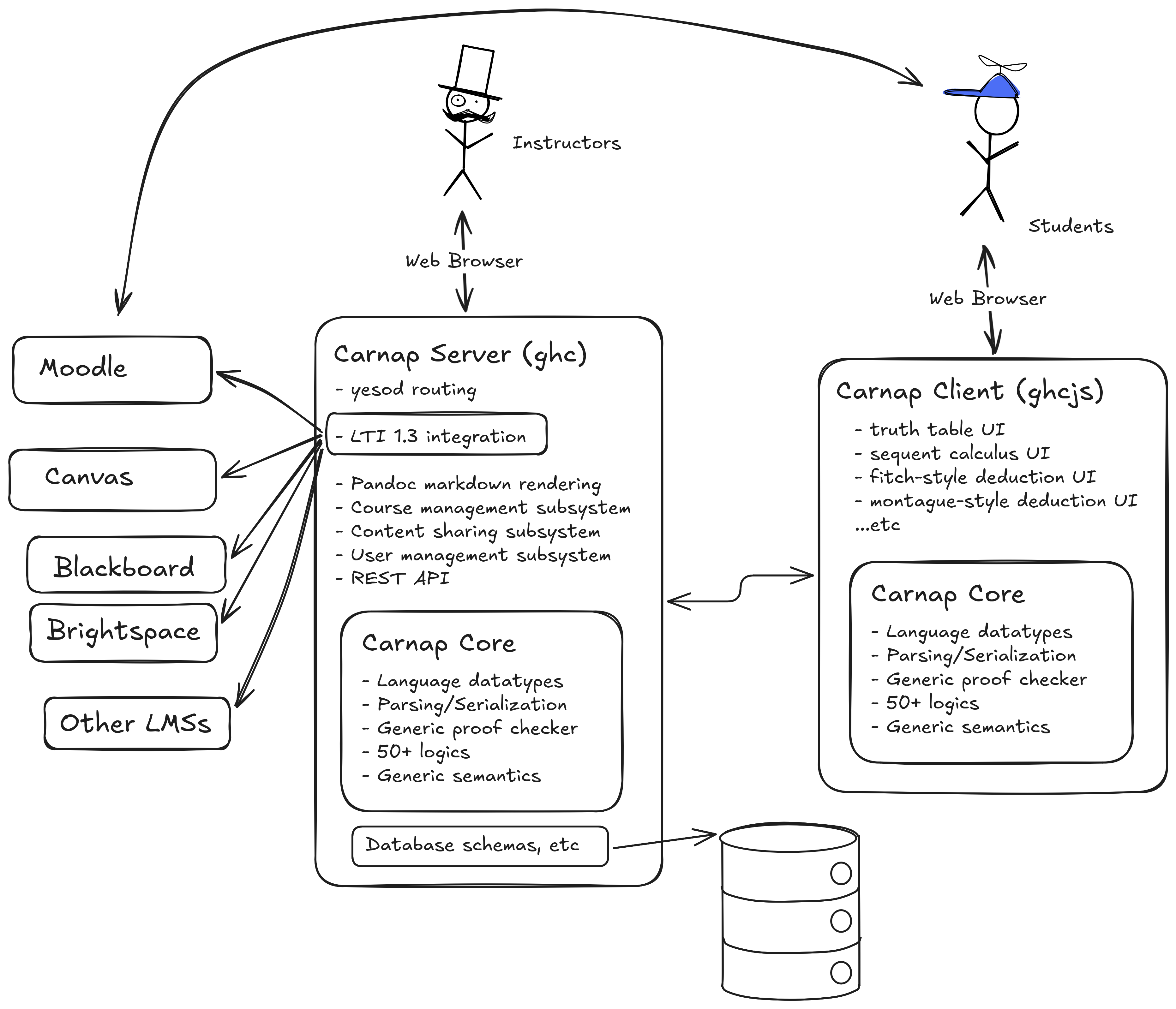}
  \caption{Architecture of Carnap.io}
  \label{fig:architecture}
\end{figure}

Figure \ref{fig:architecture} shows the architecture that the Carnap system
settled into.

The system has two main software components, demarcated more or less by the
physical boundary between the cloud server, which runs the Carnap server
application, and the primary client---a student's web browser, where the actual
proof checking takes place. The server records the student's submission and the
client's verdict, but does not reverify the proof. Both components are written
in Haskell, with the client-side code compiled to JavaScript using
\texttt{ghcjs} and the server-side code compiled to a native binary using \texttt{ghc}.

Students interact with Carnap mostly by going to the website and viewing
assignments. Each assignment is a single web page, which contains problems
interleaved with explanatory text. Depending on how the assignment is
configured, it might be closer to a textbook chapter with interactive examples,
or it might be closer to an exam. The problems include simple multiple choice
challenges, free text responses, a variety of semantic problems  and natural
language formalization challenges.\footnote{Carnap can check a student provided
formalization of an English sentence against a user-provided reference
formalization for logical equivalence, so that students can provide a
translation logically equivalent to the one the instructor suggested.} Semantic
problems cover equivalence, consistency, entailment, and joint compatibility,
assessed via truth tables or countermodel constructions. But the most
interesting and sophisticated capability by far is the proof checker: students
can write proofs in a variety of formats (the instructor chooses the appropriate
format, depending on the course), and get line-by-line feedback on the
correctness of each inference as they type.

Instructors interact with Carnap mostly through their instructor dashboard. From
the dashboard, they can perform ordinary class management tasks---assigning
partial credit, offering extensions, and creating assignments. Assignments are
uploaded as markdown documents with a special syntax for interpolating the
interactive exercises, and processed by \texttt{pandoc} (which is embedded in
the server as a library) to create the student-facing assignment web pages.
Instructors generally want to be able to assign homework that matches the
teaching materials that they've used in the past or personally developed over
the years. In order to accommodate this requirement, Carnap supports more than
fifty different formal systems, including the systems of a variety of popular
open-source and commercial textbooks. Semantic exercises generally just use the
formal system as a source of notation, controlling how formulas are parsed and
represented. But the proof checker tries to mirror the quirks of each text's
proof theory faithfully, so that it recognizes as valid all and only the proofs
that the textbook would countenance.

The whole system began its life on a Digital Ocean \$5 a month droplet, with
a single CPU and (if I recall correctly) 2GB of memory. This worked surprisingly
well for a while. Because most of the computation takes place in the client
browser, the system scales nicely to a large number of users. But eventually,
after a few scares, it became clear that there were scenarios---for example,
during an in-class exam for a class section containing several hundred
students---where the server performance could degrade under load. An exam is
paradigmatically the situation where you do \emph{not} want the server to start
misbehaving. So in order to deal with these kinds of spikes, the server was
scaled to a larger 4 CPU droplet with 8GB of total memory---still about the size
of a low-end cellphone.\footnote{
  By way of comparison, another popular browser-based education proof assistant,
  the Lean Game Server \cite{leangameserver} hosted at Heinrich Heine University Düsseldorf, is
  estimated by its developers to have a ceiling of 70 simultaneous users. That
  is because it runs proof-checking on the server rather than the client, so the
  fairly heavy load of 70 concurrent lean processes needs to live entirely on
  the server.
}

\subsection{Web-based, Free, Flexible, Engaging}

Here's what worked, and what didn't.


Making the tool web-based worked very well. It immediately solved the
early-in-the-semester deployment issues that had been such a headache with Logic
2010. It also made it straightforward to present a pleasing and uniform UI to
students, in a platform independent way. It also came with some significant
costs, mostly my fault. I wrote everything in Haskell. This made sense to me at
the time. Haskell has best-in-class support for parsing, and it let me encode
the structure of the languages in a \emph{data types \`{a} la carte} style
\cite{swierstra2008datatypes} that felt
exceptionally elegant and clear. GHCJS---a Haskell compiler that produced
JavaScript---was a young and healthy project, so I used it to compile the proof
engine and run it in the browser, thereby making possible the distributed
proof-checking approach that let me scale the system so effectively.

Drunk with power, I lost control. I wrote not only the core proof checker in
Haskell, but the entire frontend, up to and including simple things like the
truth table widgets and the multiple-choice problems. This wasn't a good idea.
There are probably some cases where Haskell makes sense for frontend work on the
web. But I don't think this was one of them.


Making the tool free (free as in beer, gratis, available without payment) also
worked very well. Students and instructors certainly appreciated it. It also
made Carnap a natural complement to the excellent ecosystem of free and open
logic teaching materials already available---in particular the family of
textbooks derived from P.D. Magnus' \emph{Forall X: An Introduction to Formal
Logic} \cite{magnusforallx} and the downstream \emph{forall x: Calgary Remix}
\cite{forallxcalgary}. Once I added support for the proof systems in those
texts, they quickly became one of the dominant modes of usage.

Making the tools free as in freedom---free and open source software---was of
more limited value. I think it was the right thing to do, morally. But it did
not have the effect I had hoped it would. The practical ideal of a free software
project is to decentralize the development load, both for robustness (in order
to keep the project alive if the main developer is hit by a bus), and to
accelerate development through parallelism. This was achieved only in a fairly
limited way, at least by the standards of a healthy open source project. A group
at the university of British Columbia contributed time and student programming
assistance, which made LTI integration possible. But generally speaking, making
the project open-source was at best a necessary condition for attracting
contributors.


Making the tool flexible worked in some ways. The double-edged sword that
I found myself wielding here was Huet's algorithm.\footnote{
  I invite the reader to refer back to the note above about being drunk with
  power.
} You're probably familiar, but this is a beautiful algorithm that can solve any
solvable higher-order unification problem.\footnote{
  Full higher order unification is undecidable, so while Huet's algorithm solves
  any solvable problem, it won't necessarily terminate if you give it an
  unsolvable unification problem---it's a semi-decision procedure only.
} The technical details (see \cite{huet1975unification}) don't really matter in the present context. It
allowed me to write the inference rules for each new proof system I wanted to
support declaratively. An inference rule just was a piece of syntax
(higher-order, meaning possibly lambda expressions) which needed to unify with
the concrete expressions provided by the user.

Huet's algorithm made it possible to quickly support new proof systems---I just
needed to turn a few knobs on the parser, depending on how a given textbook
wrote formulas, write down a few inference rules as schemata, possibly write
a small amount of special purpose code to deal with quirky eigenvariable
conditions (each Fitch-style natural deduction formalism seems to have its own
slightly different conventions), and hey presto, I had a new proof system. The
other edge of the sword, though, was that Huet's algorithm over higher order
abstract syntax was somewhat complicated to implement, and involved significant
backtracking search. Complexity begat bugs; search begat performance
bottlenecks. The fact that the algorithm made it so easy to add new proof
systems by extending the code was also probably ultimately a misfeature. It
led to a design where adding new proof systems \emph{required} touching code and
couldn't be done without having the entire build toolchain available.


The trick to the final ideal---a highly engaging proof authoring
experience---was a very tight feedback loop, closely analogous to the experience
of programming in an IDE (or using a proof assistant). When writing a proof,
each line is checked, and is either approved or rejected with minimal latency.
The student experiences a sensation of forward progress as the proof is written,
frustration when they encounter an obstruction to forward motion, and
a victorious rush of relief when they overcome that obstacle.

What I found, most of all, was that engagement requires \emph{trust}. Students
need to trust the system, and feel that any obstruction they encounter is
intrinsic to the logical problem they're facing, not a limitation of the tool.
A superficial bug, a confusing error message, a slightly inconsistent user
interface state all reduce trust in the tool. And when trust is broken,
a student's natural response is to try to replace the mental model of logic that
the instructor has been trying to impart with a new model of how the software
actually works, typically discovered through increasingly desperate rounds of
trial and error. 



\subsection{Diagnosis}

So, how did I fall short? What was missing? A glance at Figure
\ref{fig:architecture} will probably give the reader a clue. At the root of
a lot of the trouble described above is the dual-monolith architecture, based on
GHC and GHCJS. There are really two related problems: the monoliths were
promiscuously coupled systems internally, and the technologies---Haskell and
GHCJS---that felt almost uniquely well-suited for one part of the stack were
a bad fit elsewhere.

The entanglement of the various parts of the system made changes difficult.
Carnap Core's rendering and parsing were not only in the proof checker, but in
the display of formulas on the server, and the serialization of data into the
database. Any tiny change to proof checker internals needed to be checked
against the large and continuously growing library of proof systems. The
difficulty of keeping the system in your head holistically undoubtedly
discouraged open-source contributions, and it conspired with the innate
complexity of the theoretically elegant higher-order syntax and of Huet's
algorithm to slow development momentum.

The project wouldn't have been possible (not for me anyway) without GHCJS. But
GHCJS was hard to install in 2015, and it's gotten worse over time. This was
a major obstacle to onboarding contributors. I (along with some potential
contributors) experimented with various approaches to providing the toolchain
required to build Carnap. In the end, we froze everything in amber, using nix to
create a reproducible development environment and pinning a historical snapshot
of the dependencies. This worked. But asking potential contributors to install
and use nix, in order to install double-digit gigabytes of build infrastructure
was a lot to ask, especially when your potential contributors were often
professional educators rather than functional programming lunatics.

GHCJS is now deprecated, replaced by a JavaScript-emitting backend for GHC. In
principle, it would have been possible to migrate Carnap to this new backend.
But in practice (because of the promiscuous coupling mentioned above) we were
reliant on GHCJS-specific libraries for a lot of the client-side UI, and the
migration would have been painful and error-prone. Other early adopters of GHCJS
have found themselves in a similar bind. Several projects much larger than
Carnap, like \texttt{reflex-platform} \cite{reflexplatform} and Obelisk
\cite{obelisk}, still use GHCJS (managed via
nix), which ties them to GHC 8.10, a compiler version that's now more than
a half-decade behind the GHC stable release.

\subsection{Prescription}
\label{sec:prescription}

\begin{figure}
  \centering
  \includegraphics[width=.85\linewidth]{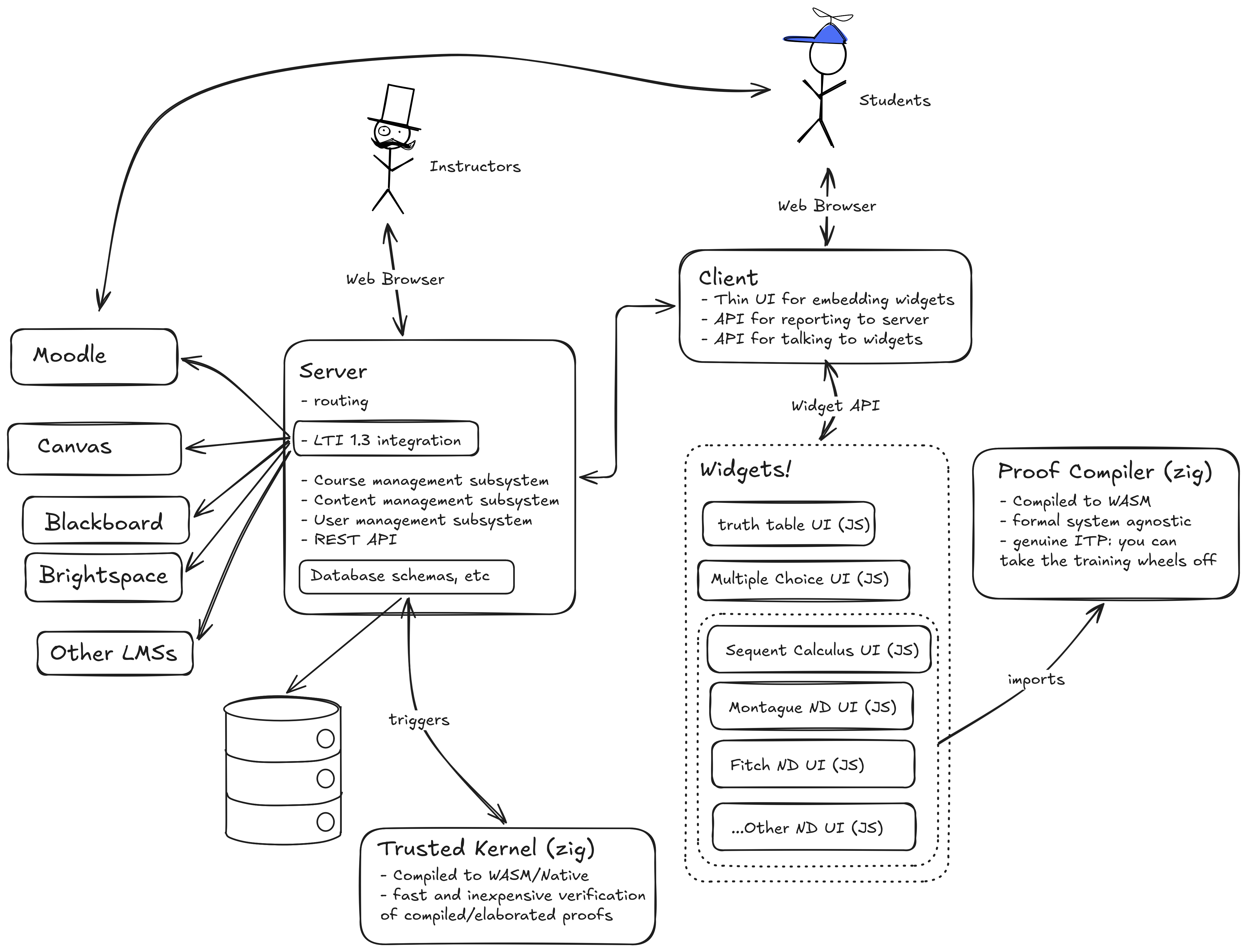}
  \caption{An Alternative Architecture}
  \label{fig:architecture-2}
\end{figure}

What would a better architecture look like? Figure \ref{fig:architecture-2}
contains a sketch. The main changes are threefold.

First, we dramatically thin the client. The new client should contain,
essentially, support for an API that makes it possible to exchange messages with
the server, and scaffolding for embedding the actual interactive components of
the system. All of the actual interactive user interface work should be ejected
into small widget libraries, decoupled from the rest of the client through
a simple widget API.

Second, we move proof-checking responsibilities (the main source of complexity
in the original client) into a separate tool, a \emph{proof compiler}. The
requirements are that 
\begin{enumerate}
  \item the compiler should run in the browser
  \item it should be formal system agnostic: not be tied to any particular logic
    system at the level of code, capable of handling new systems without code
    changes
  \item it should be a platform for genuine interactive theorem proving so
    that students have a place to go after graduating from scaffolded problem
    sets, 
  \item  it should produce efficiently auditable binary artifacts that certify
    the correctness of student work.
\end{enumerate}

Third, we reduce the server to a much more conventional CRUD system. It needs
to manage assignments, and student data, and share data with LMS systems, but
its responsibilities are strictly administrative. It should be decoupled from
the interesting parts of the code base. Proof validation (validating the
certificates generated by the proof compiler) could be handled as part of the
database integrity logic. A small trusted kernel that validates the binary
artifacts produced by the compiler can be called with each insertion, ensuring
that any student submissions that make it into the database are correct.

How do these changes address the issues described in the previous section? Well,
we've removed that promiscuous coupling that made changes and improvements
become increasingly difficult over time. Changes to the proof checking pipeline
no longer affect anything on the server or client UI. We've made it dramatically
easier for potential contributors to offer small and well-scoped additions to
the code: a new UI widget only needs to speak the widget API, it doesn't need to
know anything about the implementation details of the client or server; a new
proof system doesn't require code changes---it can be introduced by providing an
appropriate set of terms and axioms to the proof compiler.

And we can preserve the good stuff: make the system web-based, free, flexible,
and engaging.

\section{mm0-zig and abc}

The two most technically interesting components of the architecture described
above are the proof compiler, and the proof verifier. In this section, we
describe those. Both are written in zig, and can be compiled to WebAssembly in
order to run in the browser with the standard zig toolchain. The other
remaining components (the server, and the UI widgets) are still works in
progress.

\subsection{Metamath Zero}

\texttt{abc} and \texttt{mm0-zig} are both based on Mario Carneiro's Metamath
Zero (MM0), a small specification language and proof system designed around
a sharp separation between proof \emph{producers} and proof
\emph{consumers}~\cite{carneiro2020mm0}. A producer can be as elaborate as it
likes---applying tactics, performing higher-order unification, calling external
solvers---but the artifact it emits is required to be a self-contained binary
file, the MMB certificate, that a small dedicated verifier can check
independently. The underlying logic of MM0 is a multi-sorted first-order theory
with definitional extensions, expressive enough to encode systems from
propositional natural deduction up to ZFC.

The MMB format is designed for verification in essentially linear time: Carneiro
reports that MM0 currently holds the speed record for checking the entire
\texttt{set.mm} Metamath library of ZFC proofs, including 71 of Wiedijk's 100
formalization targets, at under 200 milliseconds. For our purposes, though, what
matters most is not raw speed but the size of the trusted verifier required:
small enough to fit in one person's head, and to be plausibly audited
end-to-end. The verifier can be kept manageable, and once implemented correctly,
never needs to be changed in line with the ideal of a small trusted kernel from
section~\ref{sec:prescription}.

A second consequence falls out naturally: MM0 is an ideal format for specifying
logics and languages in a way that both the verifier and the proof compiler can
understand. Adding a new proof system requires only a declaration of sorts,
function symbols, and axioms in MM0---data that \texttt{abc} consumes---rather
than a change to trusted code. The library of supported logics can grow without
enlarging the trusted base, or even the compiler.

\subsection{The Aufbau Bytecode Compiler}

The Aufbau Bytecode Compiler is a proof producer that emits checkable MMB.
It's based around a simple textual proof format that is intended both for
direct human authoring and as an intermediate representation that can be
generated from rich UIs. A proof looks like this:

\begin{lstlisting}
--- A simple hilbert system
imp_refl
--------
l1: $ a -> ((a -> a) -> a) $ by h1 
l2: $ a -> (a -> a) $ by h1 
l3: $ (a -> ((a -> a) -> a)) -> ((a -> (a -> a)) -> (a -> a)) $ by h2 
l4: $ (a -> (a -> a)) -> (a -> a) $ by mp [l1, l3]
l5: $ a -> a $ by mp [l2, l4]
\end{lstlisting}
or like this:
\begin{lstlisting}
--- A sequent calculus
not_exists_to_forall_not
------------------------
l1: $ P y ==> P y $ by ax
l2: $ P y ==> ∃ x P x $ by ex_right [l1]
l3: $ P y , ¬ (∃ x P x) ==> _ $ by not_left [l2]
l4: $ ¬ (∃ x P x) ==> ¬ P y $ by not_right [l3]
l5: $ ¬ (∃ x P x) ==> ∀ y (¬ P y) $ by all_right [l4]
\end{lstlisting}
or like this:
\begin{lstlisting}
--- natural deduction for modal logic
necessity_of_identity
---------------------
l1: $ w : x = y ⊢ w : x = y $ by ax
l2: $ w : x = y , w R v ⊢ w : x = y $ by ax
l3: $ w : x = y , w R v ⊢ v : x = x $ by eq_refl_nd
l4: $ w : x = y ⊢ w : □ (x = x) $ by box_intro [l3]
l5: $ w : x = y ⊢ w : □ (x = y) $ by eq_replace [l1, l4]
l6: $ _ ⊢ w : x = y → □ (x = y) $ by imp_intro [l5]
\end{lstlisting}
In order to handle natural deduction contexts, structural rules in the sequent
calculus, and the like, bits of syntax constructed with joiners like the comma
can be designated as checkable modulo ACUI, AU, or any structural discipline in
between. Definitions are unfolded transparently, and omitted subterms (``holes'')
can be inferred, for example in a knights and knaves
problem:
\begin{lstlisting}
alice_liar
----------
l1:  $ _ctx ⊢ says ka "¬ ka" $ by ax
l2:  $ _ctx ⊢ ka → ¬ ka $ by and_elim_l [l1]
l3:  $ _ctx ⊢ ka $ by ax
l4:  $ _ctx ⊢ ¬ ka $ by imp_elim [l2, l3]
l5:  $ _ctx ⊢ bot $ by bot_intro [l3, l4]
l6:  $ _ctx ⊢ ¬ ka $ by neg_intro [l5]
l7:  $ _ctx ⊢ ¬ ka → ka $ by and_elim_r [l1]
l8:  $ _ctx ⊢ ka $ by imp_elim [l7, l6]
l9:  $ _ctx ⊢ bot $ by bot_intro [l8, l6]
l10: $ _ ⊢ ¬ says ka "¬ ka" $ by neg_intro [l9]
\end{lstlisting}
Here, a defined term with custom syntax, like \lstinline{says A "B"} is used 
interchangeably with its definition (in this case \lstinline{A→B ∧ B→A}), and the
natural deduction context does not need to be written on each line---it can
instead be inferred as the solution to the hole \lstinline{_ctx}.

Natural deduction judgements with automatically inferred contexts were the
basic internal representation used in the original Carnap proof checker. The
flexibility of the original Carnap design thus provides some evidence that
this is a good target format for a UI to pass forward to an actual checker.

\texttt{abc} is intended to play the same role the Carnap proof checker played
internally, but with better demarcated responsibilities. Carnap used a rich
internal representation of natural-deduction judgments, higher-order
unification, and special-purpose code for idiosyncratic proof systems.
\texttt{abc} preserves the useful part of that design---the ability to
tolerate textbook-specific proof formats, implicit contexts, structural rules,
and pedagogically convenient notation---while moving the complexity into
a compiler whose output is independently auditable. And in particular, adding
a new proof system does not require changing \texttt{abc} or the trusted kernel,
but only supplying an appropriate MM0 presentation and, where necessary,
declarative annotations describing structural behavior such as associativity,
commutativity, contraction, or context inference.

\subsection{mm0-zig}

\texttt{mm0-zig} is an MM0 verifier, a proof consumer. It more or less
faithfully reimplements Mario Carneiro's original \texttt{mm0-c}, allowing us
to share some of the high-performance trusted code from the original design
between both the verifier and the compiler. It consists of about 4500 lines of
code, rounding up. Performance when compiled with zig's \texttt{ReleaseFast}
is very close to \texttt{mm0-c}, compiled with \texttt{gcc}, using
\texttt{-O2} and profile-guided optimization. Table \ref{tab:bench} shows some
performance details. Peano.mmb was generated from Peano.mm1, a reference
development of a theory of syntax within Peano arithmetic provided as part of
the MM0 repository, containing 170 terms and definitions, and 2723 axioms and
theorems.

\begin{table}[h]
  \centering
  \caption{Verification time for \texttt{peano.mmb}, measured with
    \texttt{hyperfine}. Times are wall-clock; ranges are min--max across runs.}
  \label{tab:bench}
  \begin{tabular}{lrrrr}
    \toprule
    Verifier & Mean (ms) & $\sigma$ (ms) & Range (ms) & Runs \\
    \midrule
    \texttt{mm0-c}   & 6.1 & 0.6 & 5.3--10.5 & 285 \\
    \texttt{mm0-zig} & 7.1 & 0.5 & 6.5--12.3 & 243 \\
    \bottomrule
  \end{tabular}
\end{table}

\subsection{Next steps}

Proof authoring with \texttt{abc} currently requires a text-editor or
a text-editor like interface, for example the interface currently exposed at
\url{http://grahamlk.me/Aufbau}. This is already fairly ergonomic:
\texttt{abc} provides a language-server-protocol integration, so it can
continuously check proofs in a text editor while they're being developed, and
the web editor has similar functionality.

But the raw proof text is still a bit much to ask a student to take on, at
least in a first logic class. The intention going forward is to produce richer
UIs supporting, for example, structured Fitch-style natural deduction or
Prawitz style proof trees. Those richer UIs will emit textual output
that can then be passed to an instance of \texttt{abc} for compilation. This
moves the burden of proof-checking away from frontend code, providing a clear
division of labor: widgets are responsible for providing students with pleasant
and intuitive interactions, while \texttt{abc} owns the complexity of
elaborating user input into verifiable binary proof certificates.

\section{Lessons Learned}

The reader has hopefully extracted a few portable lessons from above. But
I think there is one central point that bears repeating. Building a free and
open educational tool that will be used at scale, and long term, is
simultaneously an engineering undertaking, a social project, and a pedagogical
challenge. Engineering considerations affect long term viability (for example,
the technologies one chooses to use: will they be around in ten years?).
Design choices affect the social project (will anyone else be willing to
install your toolchain?). And implementation details affect pedagogical
effectiveness (will the system be performant, ergonomic, and stable enough to
engage students, and to earn their trust?).

I think the hardest thing to earn and maintain is trust---trust of students, of
a community of instructors, and of potential software contributors. Honesty
about the good, the bad, and the ugly of your system---and a willingness to make
changes when needed---will help earn and cement that necessary condition for
a successful project.

\bibliography{paper}

\end{document}